\documentclass[prl,twocolumn,showpacs,floatfix]{revtex4}

\usepackage{graphicx}
\begin{document}

\title{STM Imaging of Flux Line Arrangements in the Peak Effect
Regime}

\author{A. M. Troyanovski}\altaffiliation[Permanent Address : ]
{Institute for High Pressure Physics, Russian Academy of Science,
Troitsk, 142092, Russia.}\affiliation{Kamerlingh Onnes Laboratory,
  Universiteit Leiden, PO Box 9504, 2300 RA Leiden, The Netherlands}

\author{M. van Hecke}\affiliation{Kamerlingh Onnes Laboratory,
  Universiteit Leiden, PO Box 9504, 2300 RA Leiden, The Netherlands}

\author{N. Saha}\affiliation{Kamerlingh Onnes Laboratory,
  Universiteit Leiden, PO Box 9504, 2300 RA Leiden, The Netherlands}

\author{J. Aarts}\affiliation{Kamerlingh Onnes Laboratory,
  Universiteit Leiden, PO Box 9504, 2300 RA Leiden, The Netherlands}

\author{P. H. Kes}\affiliation{Kamerlingh Onnes Laboratory,
  Universiteit Leiden, PO Box 9504, 2300 RA Leiden, The Netherlands}

\date{\today}

\begin{abstract}
We present the results of a study of vortex arrangements in the
peak-effect regime of 2H-NbSe$_2$ by scanning tunneling
microscopy. By slowly increasing the temperature in a constant
magnetic field, we observed a sharp transition from collective
vortex motion to positional fluctuations of individual vortices at
the temperature which coincides with the onset of the peak effect
in ac-susceptibility. We conclude that the peak effect is a
disorder driven transition, with the pinning energy winning from
the elastic energy.
\end{abstract}


\maketitle

It is well known that the critical current density $j_c$ in weakly
disordered type II superconductors shows a sudden increase when
the applied magnetic field $H$ approaches the upper critical field
$H_{c2}$ \cite{leblanc}. An explanation for this intriguing
phenomenon, known as the peak effect, was suggested by Pippard
\cite{pippard} already in 1969. It is based on the competition
between the collective work done by the disorder (the pinning
centers) on the vortex lattice (VL) and the elastic energy stored
in vortex lattice deformations due to the pinning \cite{blatter}.
The work depends linearly on $(H_{c2}-H)$ while the elastic energy
essentially is proportional to the shear modulus $c_{66}$ of the
VL, which behaves as $(H_{c2}-H)^2$. Therefore, at some field
close to $H_{c2}$, the pinning will exceed the elasticity and the
VL will accommodate to the random pinning potential, leading to
the sudden increase in $j_c$ observed experimentally. The
discovery of the high temperature superconductors made it relevant
to also consider thermal fluctuations as a third energy scale,
which lead to melting by suppression of $c_{66}$. This would then
yield a peak effect as a precursor of the melting transition
\cite{berghuis,bhatt,ling}.

Experimentally, it is hard to probe which mechanism is
predominant, because one needs information about structural
changes of the vortex lattice positional order as a function of
time. Small angle neutron scattering (SANS) yields structural
information along the field direction averaged over the sample
volume, i.e., about the amount of entanglement of the VL. SANS has
been successfully used in the peak-effect regime
\cite{gammel,ling01,forgan} but it is resolution limited in the
transverse direction. Lorentz microscopy \cite{tonomura}, scanning
Hall probe (SHP) \cite{oral}, scanning SQUID
\cite{britton,kirtley} and, recently, magneto-optics experiments
\cite{goa} yield positional information on the scale of individual
vortices, but sense the magnetic field distribution, and only work
in the low flux density regime, usually far below $H_{c2}$, where
vortices are well separated and the vortex-vortex interaction is
very weak.
\begin{figure}
\rotatebox{0}{\includegraphics[width=8.25cm]{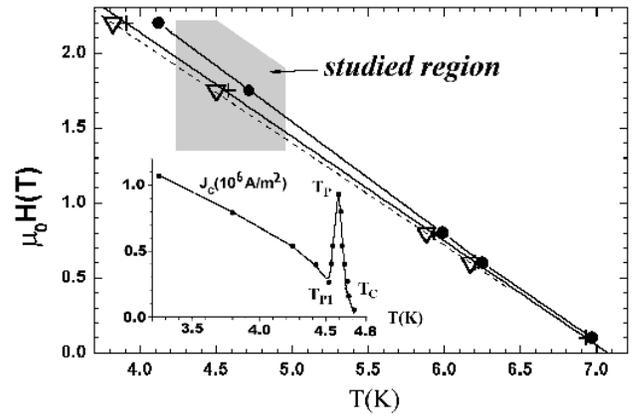}}
\caption{Phase diagram of NbSe2 obtained from susceptibility
measurements on a single crystal, showing $H_{c2}(T)$ (solid line
/ circles), the peak of critical current $H_p(T)$ (solid line /
plusses) and the onset of the peak $H_{p1}(T)$ (dashed line /
triangles). The inset shows the peak effect in the bulk critical
current density using the method due to Clem
\cite{clem94,angurel97} for a flat sample in perpendicular field
vs temperature at $\mu_0H\!=\!1.75$ T.} \label{f1-pe}
\end{figure}
In this Letter we show that information on the time-dependent
positional order can be provided by scanning tunneling microscopy
(STM), which is uniquely able to access the necessary length
scales and flux densities. We present STM data of vortex
structures obtained on a pure single crystal of 2H-NbSe$_2$ at
temperatures around 4.5~K (far below $T_c$~= 7.1~K) in the regime
of the peak effect. We observed a sharp transition from collective
vortex motion to positional fluctuations of individual vortices at
a temperature which coincides with the onset of the peak effect in
ac-susceptibility measurements. The state above the onset
temperature is characterized by the fact that the shear modulus of
the VL does not play a role and shows the properties of a pinned
liquid.

\begin{figure}[h]
\rotatebox{0}{\includegraphics[width=8.25cm]{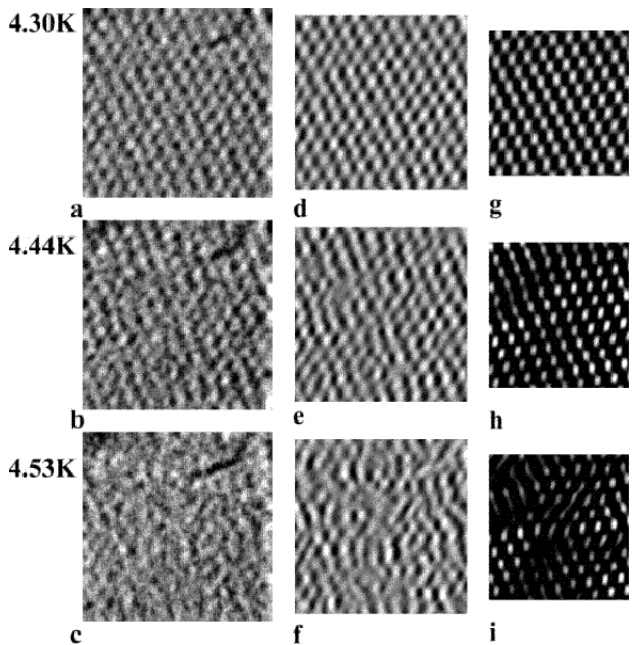}}
\caption{Raw and processed data from one sequence of STM images
acquired at 1.75 T for three different temperatures. The frame
size is about 410x460 nm.  Left (a-c) : original images (the dark
elongated stain in the upper-right corner is caused by a surface
defect); middle (d-f) : result of convolution with the shape of
"single vortex", right (g-i) : result of convolution with
"perfect" lattice cell of 3x3 units.} \label{f2-data}.
\end{figure}

Fig.~\ref{f1-pe} shows part of the $H,T$-phase diagram of one of
our crystals as obtained from magnetization measurements with the
field along the c-axis (perpendicular to the cleaved surface).
Also shown (inset) is the critical current density $j_c$ as
function of temperature $T$ in a constant field of $1.75$~T,
obtained from the maximum of the out-of-phase component of the
ac-susceptibility \cite{clem94,angurel97}. With this method the
same pinning conditions hold for the entire sample and geometrical
effects are taken into account. The peak effect sets in at $T_{p1}
\simeq$ 4.5~K, it reaches a maximum at $T_p \simeq$ 4.6~K, and
superconductivity disappears at $T_c(H) \simeq$ 4.7~K. It should
be noticed that in dc-magnetization (SQUID) measurements no peak
effect was observed due to flux creep. This means that even in the
peak-effect regime nonuniform flux distributions decay within a
few minutes to a uniform flux density.

The STM setup was the same as in a previous report
\cite{troyan99}, with the sample immersed in liquid helium. After
decreasing the magnetic field from above $H_{c2}$ at $4.27 K$ to a
preset value (e.g. 1.75~T) in persistent mode (relative decay less
than 5$\times 10^{-8}$ s$^{-1}$), we waited about 10 minutes
before starting the STM imaging. By closing the evaporation valve
during imaging a slow temperature rise was established at a rate
of 5$\times 10^{-5}$ K/s to a value of 4.9~K. The temperature was
determined from the pressure with a precision of 2~mK, but we
found a small systematic difference between the  temperatures of
the STM $T_{STM}$ and the SQUID $T_{STM}$, $T_{SQ}$,
$T_{STM}$~-$T_{SQ} \approx$ 60~mK, due to the non-uniform
temperature in the liquid. We corrected for this difference and
estimate the residual uncertainty in the temperature determination
to be about 20~mK. The bias voltage was set inside the
superconducting gap (0.2-0.8~mV) and the tunnelling current was
about 8-15~pA. The signal-to-noise (S/N) ratio goes down when
approaching $T_c(H)$, e.g., the relative conductance modulation
between vortex core and vortex cell edge decreases from 0.25 (far
below $T_{p1}$) to 0.02 (at $T_p$). Therefore we had to use a low
scanning speed of about 20-70~s per image to acquire images of
size 512 $\times$ 512 pixels. The scan area was about 500x500~nm;
at a field of 1.75~T we could probe the spatial arrangement of
about 150 vortices. The images show a small constant spatial
distortion due to a slight asymmetry in the piezo scanner. By
heating up in constant field the vortex structure is very weakly
perturbed by the almost negligible change of the equilibrium
magnetization. The effect of the tunneling current is also
expected to be negligible because it is very small ($<$~15 pA) and
it is directed parallel to the vortices. Moreover, there was no
influence on the vortex arrangement when we changed the tunnel
current from 7 to 25 pA. In this respect our experiments differ
essentially from dc and or ac-magnetization or transport
measurements which do affect the vortex configurations.

Results of experiments at 1.75 T are shown in Fig.~2a-c. We
selected three typical temperatures: (a) 4.3~K (well below
$T_{p1}$), (b) 4.44~K (just below $T_{p1}$) and (c) 4.53~K
(between $T_{p1}$ and $T_p$). The first image shows an almost
perfect vortex arrangement without large visible distortions. With
increasing temperature we see an increasing shear distortion of
the VL and the appearance of VL defects. With a further increase
of $T$ the signal from the vortices gradually decreases to the
noise level. Several image-processing procedures were used to
extract the positional information. The S/N ratio could be
considerably improved by convoluting the original data with the
images of perfect vortex lattice cells acquired at low $T$, using
cell sizes up to 10x10 vortices. In Figs.~\ref{f2-data}d-f the
results are shown of convolution with 1 vortex ('1x1'), in
Figs.~\ref{f2-data}g-i of convolution with a 3x3 cell, both with
respect to the corresponding data in the Figs.~\ref{f2-data}a-c.
Using the 1x1-convolutions, the displacements of individual
vortices could be followed up to 4.56~K, see below. The low
contrast in Figs.~\ref{f2-data}e,f is most likely due to noise,
but may also denote local vortex displacements.

\begin{figure}
\rotatebox{-0}{\includegraphics[width=8.25cm]{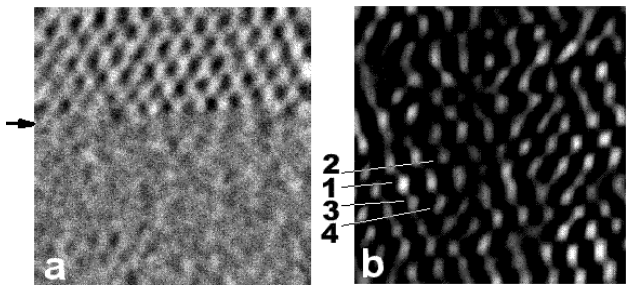}}
\caption{(a) Difference between two consecutive images at 4.31~K
in a field of 1.75~T, to demonstrate coherent motion of vortices.
When scanning the second image from bottom to top, the VL moved at
the line indicated by the arrow. (b) Average of the sum of 60
consecutive images taken with increasing temperature above
$T_{p1}$. The brightness is a measure for the probability of
finding a vortex at a certain position. The numbers indicate
individual vortices discussed in the text. } \label{f3-VLjump}
\end{figure}

With the knowledge of the vortex positions, further analysis can
be performed using the vortex motion. Inspecting consecutive
frames in the temperature interval between 4.3~K and 4.6~K in a
field of 1.75~T, we notice a sudden change in the vortex mobility
at 4.48~K. Below 4.48~K we find {\it collective} jumps of large
portions of the VL; above 4.48~K, we only see oscillations of
individual vortices around fixed positions. An example of a
collective jump is given in Fig.~\ref{f3-VLjump}a, which displays
the difference between two consecutive frames of raw data at
4.31~K. The scan direction is horizontal going from bottom to top.
In the lower part of the figure the vortex lattice did not move
substantially, and the difference gives an impression of the noise
level. At the scan position denoted by the arrow the entire VL
moved over a considerable fraction of the VL parameter $a_0$. The
next difference picture (not shown), has a lattice below and noise
above the position of the arrow. We could see no net motion of the
VL, it is jumping between metastable positions. Above 4.48~K, no
collective jumps were observed; in the 60 images in which T slowly
rose to 4.9~K (total observation time about 70 minutes) we did not
detect any net motion of individual vortices in the image,
although some vortices hopped back and forth between two
metastable positions. A movie of the change in behavior, made of
1-cell filtered frames for noise reduction, can be viewed on our
website \cite{website}.

\begin{figure}
\rotatebox{00}{\includegraphics[width=8.25cm]{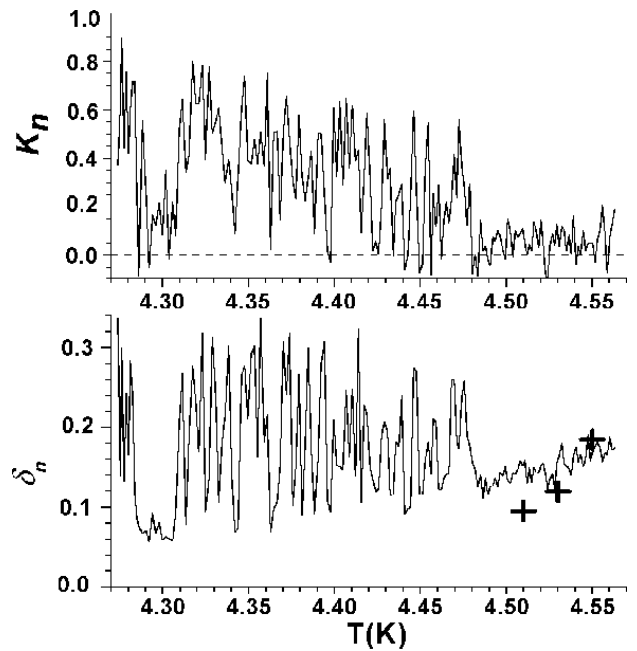}}
\caption{(a) Order parameter $K_n$ for the correlated displacement
of vortices from frame to frame as function of temperature at a
magnetic field of 1.75~T. (b) Amplitude of vortex displacement
$\delta_n$ averaged over the total number of vortices in a frame
in the same sequences of images. Symbols (+) mark the value
$\sqrt{2<u^2>}/ a_0$ (the value of $\delta_n$ for fluctuations of
individual vortices) calculated from the average of 4 selected
vortices (see Fig.~\ref{f3-VLjump} and Table~1).} \label{f4-corr}
\end{figure}

We can characterize this transition by introducing the following
two order parameters, based on the displacement $\vec{d}_{i,n}$ of
vortex $i$ going from frame $n$ to frame $n+1$, $\vec{d}_{i,n} =
\vec{r}_{i,n+1}-\vec{r}_{i,n}$; {\em{(i)}} $\delta_n$; in order to
quantify the amount of motion of the vortices, we define
\begin{equation}
\delta_n := (\frac{1}{N} \sum_i \vec{d}_{i,n}^2)^{1/2})/(a_0) \;,
\end{equation}
where the sum runs over the $N$ displacement vectors.  {\em{(ii)}}
$K_n$; in order to quantify the amount of {\em correlated} motion
of the vortices, we define
\begin{equation}
K_n := \frac{1}{N} \sum_{<i,k>}\frac{ \vec{d}_{i,n} \cdot
\vec{d}_{k,n}}{|\vec{d}_{i,n}| \cdot |\vec{d}_{k,n}|}~,
\end{equation}
where the sum runs over the displacements on neighboring vortices
$i$ and $k$ in a range 1.5 $a_0$. For both parameters, only
contributions were retained for which $\vec{d}_{i,n}$ was two or
more pixels.

Clearly, when vortices move coherently, both $\delta_n$ and $K_n$
will be large, while for uncorrelated, random motion, $K_n$ is
small but $\delta_n$ remains finite. Both $K_n$ and $\delta_n$ are
plotted versus $T$ in Fig.~\ref{f4-corr}a,b. For both, the
behavior abruptly changes at $T_{STM} \!=\!  4.48$~K, which is
essentially $T_{p1}$ (see Fig.~\ref{f1-pe}). Below $T_{p1}$ the
behavior of $K_n$ exhibits the collective vortex lattice jumps
between metastable positions in the collective pinning potential.
At $T_{p1}$, the picture suddenly changes to a situation in which
each vortex independently fluctuates within its own pin potential.
The sudden drop of $K_n$ to the noise level reflects this
transition from collective to single vortex behavior. From the
plot of $\delta_n$ versus $T$ it follows that the collective jumps
span a distance of about 0.1 - 0.25 $a_0$, which changes at
$T_{p1}$ to individual jumps with a much smaller amplitude spread.
The distance of collective jumps reflects the typical length scale
separating independent local minima in the pinning potential in
the field regime close to $H_{c2}$ (at low fields where the vortex
cores are well separated, this scale would be the Ginzburg-Landau
coherence length $\xi$). The jumps occur more frequently with
increasing temperature.

From the original STM pictures it can be also concluded that the
jumps are directed along the principal axes of the VL. In the
regime above $T_{p1}$ the mean-squared displacement gradually
grows with temperature to a value $<\!\!u^2\!\!>^{1/2} \sim
0.1$~$a_0$. This number is close to the Lindemann criterion for
melting, $<\!\!u^2\!\!>^{1/2} = c_L a_0$ with $c_L \sim
0.15-0.25$. However, we do not observe diffusive motion of
vortices over more than $a_0$. This is illustrated in Fig.~3b
where 60 images in the temperature regime above $T_{p1}$ are
averaged. Since the noise level for all vortices is supposedly the
same, the brightness of the spots represent the probability
amplitude for finding a vortex at a certain position. The
brightest spots display vortices with an almost fixed position.
One of them is marked $\#1$; other vortices (marked $\#2-4$) are
less bright, indicating larger individual positional oscillations
most likely caused by thermal fluctuations.

\begin{table}[h!]
\caption{\label{tab:table1} Mean square displacement $<u^2>^{1/2}
/ a_0$ at three temperatures between $T_{p1}$ and $T_{p}$ for the
four vortices indicated in Fig.~\ref{f3-VLjump}b. }
\begin{ruledtabular}
\begin{tabular}{lllll}
Vortex number: & $\#1$ & $\#2$ & $\#3$ & $\#4$ \\
\hline 4.51 K & 0.045& 0.095& 0.055& 0.077 \\
4.53 K& 0.055& 0.080& 0.13&0.071 \\
4.55 K&0.070& 0.10& 0.21& 0.15
\end{tabular}
\end{ruledtabular}
\end{table}

In Table 1 the values are given of $<\!\!u^2\!\!>^{1/2}/a_0$ for
these 4 vortices determined from 20 images around the denoted
temperatures. Near the onset of the peak effect (4.51~K) the
amplitudes remain far below the Lindemann criterion. Approaching
$T_p$, the amplitudes for the vortices $\#3$ and $\#4$ increase to
values close to the Lindemann criterion. However, the estimated
average of $<\!\!u^2\!\!>^{1/2}/a_0$ for all detectable vortices
coincides with the data in Fig.~\ref{f4-corr}b and remains well
below the criterion for melting. Therefore, and because of the
absence of large scale vortex diffusion within the 70 minutes
observation time, we have shown that the change at $T_{p1}$
signifies a disorder driven transition from collective pinning to
single vortex pinning. The consequence is a sudden change in the
temperature and/or field dependence of the critical current which
causes the peak effect.

Following this main result we add a few quantitative remarks about
the observed disorder and the relation of our result to those in
other papers. From the convoluted images we can extract some
information about the size of the Larkin domains in the
temperature range below the peak effect (although the limited
number of vortices in the images does not allow much statistics).
The positional order in an area of 10x10 vortices disappears at a
$T_{STM}$ of about 4.35 K, while for an area of 6x6 vortices
positional order is seen up to $T_{p1}$. The computed correlation
lengths from the Fourier transform of the images have similar
values: ~ $(4-6)a_0$ at the onset of the peak effect and $\sim
(1-2)a_0$ above the onset temperature. It should be noted this the
correlation length is about twice as large as the transverse size
of the Larkin domain $R_c$. At the onset of the peak effect the
value of $R_c/a_0$ drops from about 3 to 1. This result is in
reasonable agreement with the values we obtained in a previous
analysis of critical current data on a similar crystal
\cite{angurel97}. Based on SANS experiments on a Nb single
crystal, Ling {\it et al.} \cite{ling01} found evidence for
supercooling and superheating of the vortex matter disorder around
the peak effect line in the phase diagram and concluded that this
is direct structural evidence for a first-order vortex
solid-liquid transition. Very recent SANS work of Forgan {\it et
al.} \cite{forgan} demonstrated that the vortex solid in very pure
Nb is stable to very close to the $H_{c2}(T)$ line. This shows
that the history dependence observed by Ling {\it et al.} is
disorder driven rather than of thermal origin and this is in
agreement with the disorder driven transition seen in our STM
experiment.

In conclusion, the mechanism for the peak effect is the conquest
of the pinning energy over the elastic energy, just as was first
suggested by Pippard \cite{pippard}, and later in more detail
described by Larkin and Ovchinnikov \cite{Larkin79}. The high
temperature phase is characterized by the fact that the shear
energy has become irrelevant. In this phase, the vortices are not
free to move around as one would expect to occur in a liquid, but
they remain trapped in their own pin potential which may differ
for each vortex. In terms of the collective pinning theory this
would be the amorphous limit, but it would be equally correct to
say that we are dealing with a pinned liquid.

This work is part of the research program of the "Stichting FOM",
which is financially supported by NWO. It was partially supported
by the Dutch-Russian science collaboration financed by NWO, and by
the ESF network 'Vortex'. We thank R. Drost for assistance in the
early experiments.


\begin{thebibliography}{99}

\small

\bibitem{leblanc} M. A. R. Le Blanc and W. A. Little , in {\em Proceeding
of the VII International Conference on Low Temperature Physics
1960} (University of Toronto Press, Toronto), p.198, 1960.

\bibitem{pippard} A. B. Pippard, Philos. Mag. {\bf 19}, 217 (1969).

\bibitem{blatter} G. Blatter {\em et al.}, Rev. Mod. Phys. {\bf66}, 1125
(1994).

\bibitem{berghuis} P. Berghuis and P.H. Kes, Phys. Rev. B {\bf47}, 262 (1993).

\bibitem{bhatt} S. Bhattacharya and M.J. Higgins, Physica C {\bf 257},
232 (1996).

\bibitem{ling} X.S. Ling, J.E. Berger and D.E. Prober, Phys. Rev. B
{\bf 57}, R3249 (1998).

\bibitem{gammel} P.L. Gammel {\em et al.}, Phys. Rev. Lett. {\bf80}, 833
(1998).

\bibitem{ling01} X. S. Ling {\em et al.}, Phys. Rev. Lett. {\bf86}, 712
(2001).

\bibitem{forgan} E.M. Forgan {\em et al.}, Phys. Rev. Lett. {\bf 88}, 167003
(2002)

\bibitem{tonomura} A. Tonomura {\em et al.}, Nature, {\bf412}, 620 (2001).

\bibitem{oral} A. Oral {\em et al.}, Phys. Rev. Lett. {\bf 80}, 3610 (1998).

\bibitem{britton} B. L. T. Plourde {\em et al.}, Physica C {\bf 341},
1023 (2000).

\bibitem{kirtley} J. R. Kirtley {\em et al.}, Phys. Rev. Lett. {\bf 76},
1336 (1996).

\bibitem{goa} P. E. Goa {\em et al.}, Supercond. Sci. Technol. {\bf 14},
729 (2001).

\bibitem{marchevsky} M. Marchevsky, M. J. Higgins and S. Bhattacharya,
Nature, {\bf 409} 591 (2001).

\bibitem{clem94} J. R. Clem and A. Sanchez, Phys. Rev. B 50, 9355 (1994).

\bibitem{angurel97} L.A. Angurel {\em et al.}, Phys. Rev. B 56, 3425 (1997).

\bibitem{troyan99} A.M. Troyanovski, J.Aarts, P.H.Kes, Nature, 399, 665
(1999).

\bibitem{website} http://www.physics.leidenuniv.nl/sections/cm/msm

\bibitem{Larkin79} Larkin A. I. and Ovchinnikov Yu. N., J. Low
Temp. Phys. 34, 409 (1979).

\end{thebibliography}
\end{document}